\documentclass{iaus}
\usepackage{graphicx}

\title[Gas flows in galaxies.] 
{Gas flows in galaxies: the relative importance of mergers and bars.}

\author[Ellison et al.]   
{Sara L. Ellison$^1$,
David R. Patton$^2$,
Preethi Nair$^3$,
Luc Simard$^4$,
J. Trevor Mendel$^1$,
Alan W. McConnachie$^4$,
Jillian M. Scudder$^1$}

\affiliation{$^1$Department of Physics and Astronomy, University of Victoria, Victoria, British Columbia, V8P 1A1, Canada.  \\ 
$^2$ Department of Physics \& Astronomy, Trent University, 
1600 West Bank Drive, Peterborough, Ontario, K9J 7B8, Canada.\\
$^3$ INAF-Astronomical Observatory of Bologna,
Via Ranzani 1, 40127 Bologna, Italy.\\
$^4$ National Research Council of Canada,
Herzberg Institute of Astrophysics, 5071 West
Saanich Road, Victoria, British Columbia, V9E 2E7, Canada}

\pubyear{2011}
\volume{277}  
\jname{Tracing the Ancestry of Galaxies on the Land of our Ancestors}
\editors{Carignan, C., Freeman, K. C. \& Combes, F.  eds.}
\begin{document}

\maketitle

\begin{abstract}
Galaxy-galaxy interactions and large scale galaxy bars are usually
considered as the two main mechanisms for driving gas to the centres
of galaxies.  By using large samples of galaxy pairs and visually
classified bars from the Sloan Digital Sky Survey (SDSS), we compare
the relative efficiency of gas inflows from these two processes.  We
use two indicators of gas inflow: star formation rate (SFR) and gas
phase metallicity, which are both measured relative to control samples.
Whereas the metallicity of galaxy pairs is suppressed relative to its
control sample of isolated galaxies, galaxies with bars are metal-rich
for their stellar mass by 0.06 dex over all stellar masses.  The SFRs
of both the close galaxy pairs and the barred galaxies are enhanced by
$\sim$ 60\%, but in the bars the enhancement is only seen at stellar
masses M$_{\star} >$ 10$^{10}$ M$_{\odot}$. Taking into account the
relative frequency of bars and pairs, we estimate that at least three
times more central star formation is triggered by bars than
by interactions.

\end{abstract}

\section{Introduction}

The relative importance of external versus internal processes in
galaxy evolution is an ongoing debate in astronomy.  One contemporary
example of this debate was discussed by Chris Conselice at this
meeting, namely the question of whether galaxy mass (an intrinsic
property) or environment (external influence) has a larger impact on a
galaxy's evolution.  In this contribution, we will investigate what
mechanism is most important for triggering gas flows to galactic
centres, considering both the internal process of bar formation and
the external effect of galaxy-galaxy interactions.  Both of these
processes have been well-documented in the literature to trigger star
formation due to the inflow of gas through tidal torques and angular
momentum loss (e.g. Martinet \& Friedli 1997;
Barton et al. 2000; Nikolic et al. 2004; see also
the contributions by Perez and Di Matteo in these proceedings).
However, a direct comparison of the two mechanisms requires a
large, homogeneous dataset and consistent technical analyses.

We note that bars can themselves be formed during interactions, so the
two processes are not entirely independent.  However, only 4 of the
bars in our sample appear to be currently undergoing a strong
interaction, consistent with typical merger fractions at this low
redshift.  We therefore consider that bars represent a much more
extended phase in the galaxy's history than a fly-by or merger.  Our
comparison is therefore also one of timescales, comparing the effect
of a close encounter which is short-lived but potentially dramatic,
and the longer-lived bar phase (which may be either
interaction-induced or secular).

\section{Sample and Analysis}

We have selected a sample of spectroscopic galaxy pairs from the SDSS
DR7 with small velocity differences ($\Delta V < 300$ km/s) and projected
separations ($r_p <$ 30 kpc).  Similarly, we have used a sample of visually
classified bars from the SDSS DR4 compiled by Nair \& Abraham (2010a).  We
further require that reliable spectroscopic SFRs and gas phase metallicities
are available, resulting in samples of bars and pairs that contain 311
and 431 galaxies respectively.  Details of the full sample selection
can be found in Ellison et al. (2008a, 2010, 2011) and Patton et al. (2011).
An important component of our analysis is the construction of control samples
matched simultaneously in stellar mass and redshift from a pool of galaxies
without close companions (in the case of the galaxy pair control sample)
or unbarred galaxies (in the case of the barred galaxy sample).  
The matching is done
iteratively and without replacement until a Kolmogorov-Smirnov test drops below
30\% for either a comparison of the masses or redshifts.  Matching multiple
control galaxies to each test galaxy greatly reduces the statistical
uncertainties in the properties of the control sample.

To quantify the effect of gas inflows, we use two metrics: SFR and metallicity.
Montuori et al. (2010) have shown that gas inflows simultaneously trigger
central star formation and result in an initial dilution of the gas phase
metallicity.  After the starburst is complete, the galactic interstellar
medium (ISM) is gradually enriched by the nucleosynthetic products of the
triggered star formation.  Studying SFRs and metallicity therefore not
only provides evidence of gas inflows, but also a timescale on which these
processes occur.  To quantify changes in the SFR and metallicity, the
mass-SFR and mass-metallicity relations of the bar/pair control samples
are fit, such that we can predict the expected SFR and metallicity of a galaxy
at a given stellar mass.  These fits are performed on the fibre quantities
which, in combination with the redshift matching, mitigates the impact of
aperture effects (the SDSS fibres cover only the inner few
kpc of the galaxies).  The actual values of SFR and metallicity in the 
bars/pairs are compared with the predicted values (for their stellar mass)
and an offset ($\Delta$) is calculated from the difference.

\section{Results: Galaxy Pairs}

\begin{figure}[b]
\begin{center}
 \includegraphics[width=5.4in]{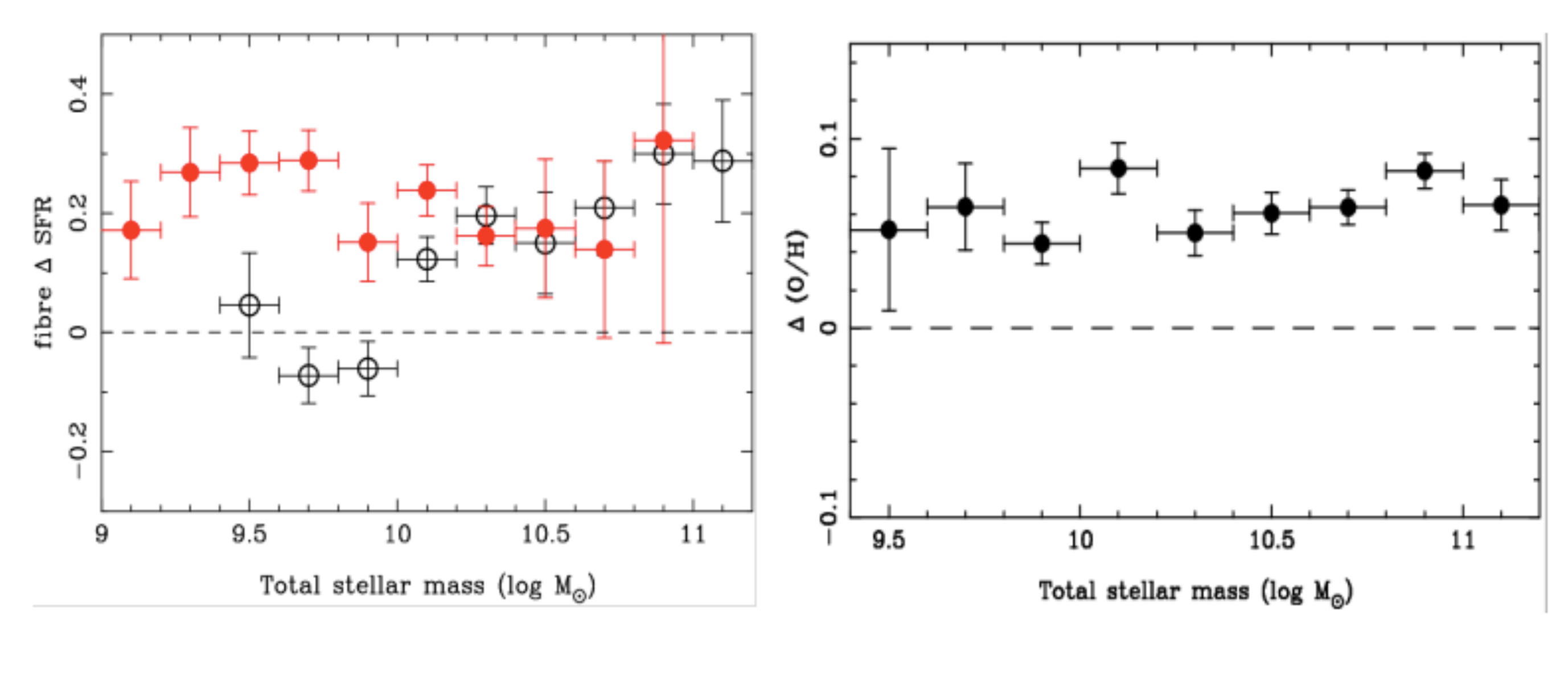} 
 \caption{Left panel: the enhancement in fibre star formation rate ($\Delta$
SFR) as a function of stellar mass for bars and pairs.  Enhancements are
relative to samples matched in mass and redshift.  Solid
points show the enhancement in close pairs with $\Delta V <$ 300 km/s and
$r_p < 30$ kpc.  Open points show the enhancement for barred galaxies
in the visually classified sample of Nair \& Abraham (2010a).  Right
panel:  enhancements in metallicity in barred galaxies relative to a
control sample of unbarred galaxies.}
   \label{fig1}
\end{center}
\end{figure}

Previous studies of close galaxy pairs have shown that galaxy
interactions result in low metallicities for their luminosity (Kewley et
al. 2006) and high SFRs (e.g. Barton et al. 2000; Nikolic et al. 2004,
amongst many others).  However, the large samples afforded by SDSS
allow us to tease apart the dependences of these offsets as a function
of higher order properties.  These results have already been published
in the literature and we only briefly review the results of our group.

Considering first the metallicity.  Ellison et al. (2008a) showed that
about 50\% of the offset in the luminosity-metallicity relation in pairs
is due to increased luminosity.  Considering the mass-metallicity relation,
it was shown that pairs are metal-poor for their mass by only 0.03 dex,
supporting the interpretation that interactions are experiencing metallicity
dilution due to gas inflows.

Turning now to SFRs, Ellison et al. (2008a) showed that triggered SFRs
are highest in the major (more equal mass) interactions.  Furthermore,
Ellison et al.  (2010) demonstrated that the triggered SFR depends
strongly on environment.  Galaxy pairs in low density environments are
enhanced by 60\%, but no SFR enhancement is seen in pairs in the
highest densities.  Asymmetries in galaxy morphologies exist in the
closest separation pairs in all environments, indicating that although
interactions occur over a wide range of environments, star formation
is triggered preferentially at low densities.  This is likely due to
the prevalence of gas-rich galaxies in low density environments, which
have a ready supply of fuel for star formation.  Finally, the
triggered star formation is central as evidenced by much bluer colours
in the bulge (Ellison et al. 2010) and fibre (Patton et al. 2011)
colours.  Global colours are much less affected and disks do not
show any colour change at close separations.

\section{Results: Galaxy Bars}

Like galaxy pairs, barred galaxies show an increase in their SFRs (at
a given stellar mass) relative to the unbarred control sample by
$\sim$ 60\%.  However, as shown in the left panel of Figure 1, whereas 
the pairs' SFR
enhancement is seen at all stellar masses, in barred galaxies it is
only seen for M$_{\star} >$ 10$^{10}$ M$_{\odot}$.  This mass threshold
has been shown by Nair \& Abraham (2010b) to correspond to the fairly
rapid transition between the weakly barred, low mass, late-type
spirals and the more strongly barred, higher mass early types.

In contrast to galaxy pairs, barred galaxies show an enhanced
metallicity for a given stellar mass by $\sim$ 0.06 dex (Figure 1, right
panel).  This
indicates that bars are relatively long-lived and the star formation
is likely to be extended in time with multiple bursts (see also
contributions by Perez and Robert in these proceedings).
Interestingly, the metal enhancement is seen even at M$_{\star} <$
10$^{10}$ M$_{\odot}$ where there is no enhanced star formation.  This can be
explained by an early, but short-lived, period of enhanced star
formation at low mass, such that we see the chemical enrichment, but the
actual starburst is long passed.  This interpretation is supported
by the models of Combes \& Elmegreen (1993) who find that high mass
galaxies are able to grow their bars over a longer period of time, and
to greater extents than lower mass galaxies.

It has been recently demonstrated that the mass-metallicity
relation for the general star-forming galaxy population
is itself modulated by SFR, such that galaxies with higher
SFRs tend to have lower metallicities (Ellison et al. 2008b).
Mannucci et al. (2010) and Lara-Lopez et al. (2010) have even
suggested a fundamental relation between SFR, mass and metallicity
in star-forming galaxies that can be fit with a plane.  Interestingly,
barred galaxies do not follow this general trend, since they have
both enhanced SFRs and higher metallicities for their mass, and would
therefore presumably be outliers on the `fundamental relation'.

Finally, to compare the relative impact of bars and interactions on
triggered star formation ($\epsilon_{b/p}$) we must consider the
relative fraction of bars and pairs in the parent galaxy sample
($f_b/f_p$) and the fraction of bars and pairs that made it into our
star-forming (emission line selected) sample ($f_{b,\star}/f_{p,\star}$), i.e.

\begin{equation}\label{c_eqn}
\epsilon_{b/p} = \frac{f_{b}}{f_{p}} \times \frac{f_{b,\star}}{f_{p,\star}} \times \frac{10^{\Delta SFR_b}}{10^{\Delta SFR_p}}.
\end{equation}

We find $\epsilon_{b/p} \sim 3$.  However, this is likely to be a lower
limit since visually classified barred samples in the optical tend to
yield relatively low bar fractions (e.g. relative to the IR).  We therefore
conclude that bars contribute at least 3 times more to the centrally
triggered star formation than interactions.  For gas flows, internal
processes seem to outweigh external mechanisms.





\end{document}